\documentclass[twocolumn]{emulateapj}
\shorttitle{Spin-Orbit Alignment}
\shortauthors{Fabrycky \& Winn}
\usepackage{natbib}

\begin{document}

%
\def\no{\mathbf{n}_o}
\def\ns{\mathbf{n}_s}
\def\lsim{\lower.5ex\hbox{\ltsima}}
\def\gsim{\lower.5ex\hbox{\gtsima}}
\def\gtsima{$\; \buildrel > \over \sim \;$}
\def\ltsima{$\; \buildrel < \over \sim \;$}

\title{Exoplanetary Spin-Orbit Alignment:\\
Results from the Ensemble of Rossiter-McLaughlin Observations}

\slugcomment{ApJ: submitted Dec. 19, 2008; accepted Feb. 3, 2009}

\author{
Daniel C.~Fabrycky\altaffilmark{1,2},
Joshua N.~Winn\altaffilmark{3}
}

\altaffiltext{1}{Harvard-Smithsonian Center for Astrophysics, 60
  Garden St, MS-51, Cambridge, MA 02138; {\tt
    daniel.fabrycky@gmail.com}}

\altaffiltext{2}{Michelson Fellow}

\altaffiltext{3}{Department of Physics, and Kavli Institute for
  Astrophysics and Space Research, Massachusetts Institute of
  Technology, Cambridge, MA 02139}

\begin{abstract}

  One possible diagnostic of planet formation, orbital migration, and
  tidal evolution is the angle $\psi$ between a planet's orbital axis
  and the spin axis of its parent star. In general, $\psi$ cannot be
  measured, but for transiting planets one can measure the angle
  $\lambda$ between the sky projections of the two axes via the
  Rossiter-McLaughlin effect. Here, we show how to combine
  measurements of $\lambda$ in different systems to derive statistical
  constraints on $\psi$. We apply the method to 11 published
  measurements of $\lambda$, using two different single-parameter
  distributions to describe the ensemble. First, assuming a Rayleigh
  distribution (or more precisely, a Fisher distribution on a sphere),
  we find that the peak value is less than
  $22^\circ$ with 95\% confidence. Second, assuming a fraction $f$ of
  the orbits have random orientations relative to the stars, and the
  remaining fraction ($1-f$) are perfectly aligned, we find $f < 0.36$
  with 95\% confidence. This latter model fits the data better than
  the Rayleigh distribution, mainly because the XO-3 system was found
  to be strongly misaligned while the other 10 systems are consistent 
  with perfect alignment.  If the XO-3 result proves robust, then our 
  results may be interpreted as evidence for two distinct modes of 
  planet migration.

\end{abstract}


\section{Introduction}

In a planetary system, the angle $\psi$ between the orbital axis and
the stellar rotation axis may provide clues about the processes that
sculpt planetary orbits. As an example, the angle between the Sun's
rotation axis and the north ecliptic pole is $\psi_\odot = 7.15$~deg
(see, e.g., \citealt{2005BG}). The smallness of $\psi_\odot$, along
with the small mutual inclinations between planetary orbits, is prima
facie evidence for formation in a spinning disk. The smallness of
$\psi_\odot$ has also been used to constrain the properties of any
``Planet X'' or solar companion star \citep{1972GW}, and to place
upper bounds on violations of Lorentz invariance \citep{1987N}. That
$\psi_\odot$ is not even closer to zero has been interpreted as
evidence for an early close encounter with another star \citep{1993H}
or a non-axisymmetric, ``twisting'' collapse of the Sun's parent
molecular cloud \citep{1991T}.

For exoplanets, it has been recognized that $\psi$ is a possible
diagnostic of theories of planet migration. Some of the mechanisms
that have been proposed to produce close-in giant planets would
preserve an initial spin-orbit alignment \citep{1996L,1997W,
  1998Murray}, while others would produce at least occasionally large
misalignments \citep{2001F, 2001YT, 2001PT, 2002TP, 2002MW, 2003TL, 2007WMR,
2007FT,2007C, 2007JT, 2008Nagasawa}.  Tides raised on the star are not expected to play a
major role in altering $\psi$ \citep{2005Wa}, but it is
possible that coplanarization is more efficient than expected
\citep{2008M, 2008P}. For example, if a hot Jupiter migrated inward before
its host star contracted onto the main sequence, the distended stellar
envelope could produce more pronounced tidal effects
\citep{1989ZB,2004DLM}.  

Independent of the interpretation, the angle $\psi$ is a fundamental
geometric property, and for this reason alone it is worth seeking
empirical constraints on $\psi$ whenever possible. We regard $\psi$ to
be on a par with the semimajor axis and the eccentricity: all of them
are basic orbital parameters for which accurate and systematic
measurements can lead to revealing discoveries and statistical
constraints on exoplanetary system architectures.

For a generic exoplanet discovered by the Doppler method, no
information about spin-orbit alignment is available. For transiting
exoplanets, one may exploit a spectroscopic phenomenon known as the
Rossiter-McLaughlin (RM) effect. During a transit, the planet hides
part of the rotating stellar disk and causes the stellar spectral
lines to be slightly distorted. The distortion is usually manifested
as an ``anomalous'' Doppler shift of order $\Delta V = -(R_p/R_s)^2
V_p$, where $R_p/R_s$ is the planet-to-star radius ratio, and $V_p$ is
the projected rotation rate of the hidden portion of the stellar
photosphere \citep{2005O, 2006G, 2007GW}. Because photometric
observations give a precise and independent measure of $(R_p/R_s)^2$,
spectroscopic monitoring of $\Delta V$ reveals $V_p(t)$, thereby
allowing one to chart the planet's trajectory relative to the
sky-projected stellar rotation axis.

An important limitation of the RM technique is that it is sensitive
only to the angle $\lambda$ between the {\it sky projections}\, of the
orbital and rotational axes.\footnote{Strictly speaking, the RM signal
  depends more directly on the angle $\lambda'$ between the transit
  chord and the sky-projected stellar rotation axis. For an eccentric
  orbit this angle may differ from the angle $\lambda$ between the sky
  projections of the orbital axis and the stellar rotation axis. It is
  straightforward to relate $\lambda$ to $\lambda'$ when the orbital
  eccentricity and argument of pericenter are known from the Doppler
  orbit of the star. For the systems considered in this paper, the
  maximum difference between $\lambda$ and $\lambda'$ is approximately
  $2\arcdeg$ (for HAT-P-2) and is in all cases much smaller than the
  measurement uncertainty in $\lambda$.} We refer to $\psi$ as the
spin-orbit angle, and to $\lambda$ as the projected spin-orbit
angle. In general the line-of-sight component of the stellar rotation
axis is unknown. When $|\lambda|$ is small, then $|\lambda|$ is a
lower limit on $\psi$. (The situation is a bit more complex for large
$|\lambda|$, as will be shown in this paper.) While the finding of a
large value of $\lambda$ implies that there is a large spin-orbit
misalignment, with consequent implications for the system's dynamical
history, the finding of a small value of $\lambda$ has a more
ambiguous interpretation.

The way to overcome this limitation and draw general inferences about
spin-orbit alignment is to consider the ensemble of RM results. The
situation is similar to the early days of Doppler planet
detection. Doppler measurements give only $M_p \sin i_o$, where $M_p$
is the planet's mass and $i_o$ is the orbital inclination. When there
were only a few detections, it was impossible to draw firm conclusions
about the mass distribution of the planets, or even to be completely
certain that they were planets and not brown dwarfs in face-on orbits
\citep{1998M,2000SB}. However, once tens of systems were known with
precise measurements of $M_p \sin i_o$, the planetary mass
distribution came into focus \citep{2001J,2001ZM,2002TT}, under the
reasonable assumption that the orbits are randomly oriented in space.

There are now 11 exoplanetary systems for which RM measurements have
been reported. The results are summarized in Table~\ref{table:rm}. The
time is ripe to undertake an analogous study of the statistical
constraints on spin-orbit alignment.  It is worth drawing attention to the entries for HD~209458
and HD~149026, for which we are using updated determinations of
$\lambda$ by Winn \& Johnson~(in preparation). For HD~209458, the
revision is due to an improved analysis method taking into account
correlated errors in the radial-velocity data. For HD~149026, a better
transit light curve led to enhanced precision in $\lambda$.

\begin{deluxetable}{lcc}
\tablecaption{Summary of RM measurements\label{table:rm}}
\tablewidth{0pt}

\tablehead{
\colhead{Exoplanet} &
\colhead{Projected spin-orbit angle $\lambda$~[deg]} &
\colhead{References}
}

\startdata
HD~189733b   &$-1.4 \pm 1.1$ & 1 \\
HD~209458b   &$0.1 \pm 2.4$ & 2,3,4,5,6$^\star$ \\
HAT-P-1b     &$3.7 \pm  2.1$ & 7  \\
CoRoT-Exo-2b &$7.2 \pm  4.5$ & 8  \\
HD~149026b   &$1.9 \pm 6.1$ & 9,6$^\star$ \\
HD~17156b    &$9.4 \pm  9.3$ & 10,11$^\star$ \\
TrES-2b      &$-9.0\pm 12.0$ & 12 \\
HAT-P-2b     &$1.2 \pm 13.4$ & 13$^\star$,14 \\
XO-3b        &$70.0\pm 15.0$ & 15 \\
WASP-14b     &$-14.0\pm 17.0$ & 16 \\
TrES-1b      &$30.0 \pm 21.0$ & 17
\enddata
\tablecomments{References: (1)
  \cite{2006W}; (2) \cite{2000Q}; (3) \cite{2000BM}; (4)
  \cite{2005Wittenmyer}; (5) \cite{2005Wa}; (6) Winn \& Johnson
  (in prep.); (7) \cite{2008J}; (8) \cite{2008Bo}; (9) \cite{2007Wo};
  (10) \cite{2008N}; (11) \cite{2008C}; (12) \cite{2008Wa}; (13)
  \cite{2007Wb}; (14) \cite{2008L}; (15) \cite{2008H}; (16)
  \cite{2008Joshi}; (17) \cite{2007N}.
  Where multiple references are given, the quoted
  result is taken from the starred reference. 
}

\end{deluxetable}

The results in Table~1 are easily summarized: each individual system
besides XO-3 is consistent with perfect spin-orbit alignment within
2$\sigma$. However it is not obvious what exactly is ruled out by
these results, or what we may conclude about the ``typical'' value of
$\psi$ among the transiting planets. The purpose of this paper is to
provide a statistical framework for understanding statistical
constraints on spin-orbit alignment that follow from RM observations,
and apply it to the current data. We are primarily concerned with the
empirical information about the distribution of $\psi$, rather than
the interpretation in terms of migration theories or tidal effects,
which will be the subject of future studies.

This paper is organized as follows. The geometry of this problem is
defined in \S~\ref{sec:geometry}. The relevant probability
distributions for individual systems are derived in
\S~\ref{sec:probind}. Constraints on $\psi$ based on an ensemble of RM
observations are discussed in
\S~\ref{sec:ensemble}. \S~\ref{sec:discussion} gives a summary of the
results, a discussion of some limitations of our analysis, and some
suggestions for future work.

\section{Spherical Geometry of the Rossiter-McLaughlin Effect}
\label{sec:geometry}

Let $\no$ and $\ns$ denote the unit vectors in the directions of the
orbital angular momentum and stellar rotational angular momentum,
respectively. The angle between $\no$ and $\ns$, is the ``spin-orbit
angle,'' denoted $\psi$. This is presumably the only angle of
intrinsic physical significance in this problem, possibly bearing
information about the initial condition for planet formation, the
endpoint of planet migration, or the result of tidal
evolution. However, $\psi$ is not directly measurable, and we must
introduce some other angles.

Figure~1 shows two useful coordinate systems. In the
``observer-oriented'' coordinate system shown in the left panel,
$\hat{Z}$ points at the observer, $\hat{X}$ points along the line of
nodes of the planetary orbit, and $\hat{Y}$ completes a right-handed
triad. The ascending node of the planet (the location where the planet
pierces the sky plane with $\dot{Z}>0$) is at $X<0$. In this
coordinate system, $\no$ is in the $YZ$ plane and is specified by the
inclination angle $i_o = \arccos(\no\cdot\hat{Z})$, which ranges from
$0$ to $\pi$. Specifying $\ns$ requires two angles, the inclination
angle $i_s = \arccos(\ns\cdot\hat{Z})$ and an azimuthal angle, which
by the convention of Ohta, Taruya, \& Suto (2005) we take to be
$\lambda$, measured clockwise on the sky from the $Y$-axis to the sky
projection of $\ns$. In summary,
\begin{eqnarray}
{\mathbf n}_o & = & \hat{Y} \sin i_o + \hat{Z} \cos i_o \\
{\mathbf n}_s & = & \hat{X} \sin i_s \sin \lambda +
                   \hat{Y} \sin i_s \cos \lambda +
                   \hat{Z} \cos i_s . \label{eq:ns}
\end{eqnarray}

\begin{figure*}[ht]
\epsscale{0.85}
\plotone{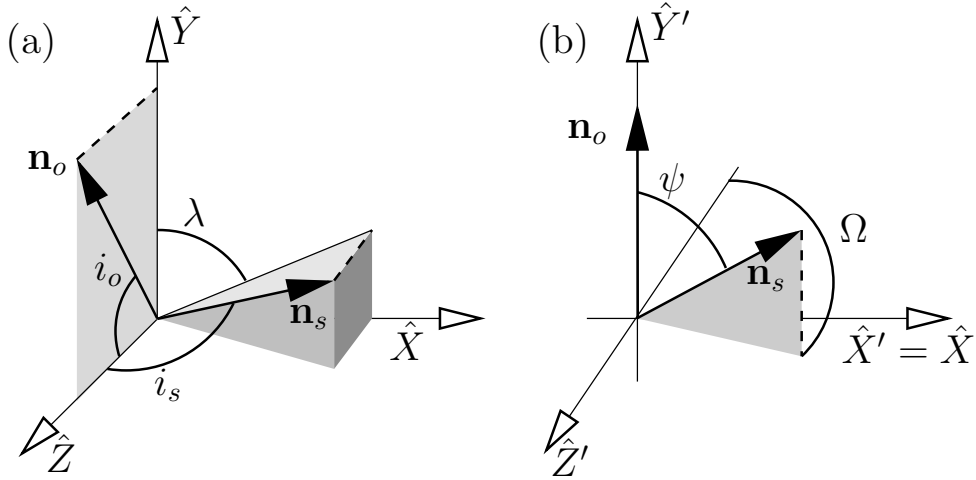}
\caption{ Two useful coordinate systems. (a) An ``observer-oriented''
  coordinate system, in which $\hat{Z}$ points toward the observer and
  the $X$--$Y$ plane is the sky plane. (b) An ``orbit-oriented''
  coordinate system in which the $\hat{Y}'$ axis is the orbital axis,
  and the $\hat{X}'$--$\hat{Z}'$ plane is the orbital plane. The two
  coordinate systems are related by a rotation of $\pi/2 - i_o$ about
  the $\hat{X}$=$\hat{X}'$ axis.}
\label{fig:coords}
\vspace{0.3 in}
\end{figure*}

For a transiting planet, $i_o$ is measurable via transit photometry,
$\lambda$ (the ``projected spin-orbit angle'') is measurable via the
RM effect. Usually there is no direct measurement of $i_s$, although
it is possible to constrain $i_s$ via asteroseismology \citep{2003GS} or by combining estimates of the stellar radius, stellar
rotation period, and projected rotational velocity (see, e.g., \citealt{2007Wa}). 
By symmetry, a configuration ($i_o$,$\lambda$) cannot be
distinguished from a different configuration $(\pi-i_o,
-\lambda)$. Because of this degeneracy we restrict $i_o$ to the range
$[0,\pi/2]$, and allow $\lambda$ to range from $-\pi$ to $+\pi$. A
positive (negative) value for $\lambda$ means that, from the observer's
perspective, the projected stellar spin axis is rotated clockwise 
(counterclockwise) with respect to the projected orbit normal.
Values of $|\lambda|$ greater than $\pi-i_o$ correspond to
retrograde orbits.

In the ``orbit-oriented'' coordinate system $X'Y'Z'$ shown in the
right panel of Figure~1, we define $\hat{Y'}\equiv\no$. This system is
related to $XYZ$ by a rotation of $\pi/2 - i_o$ about the $X$ axis. We define
$\psi$ and $\Omega$ as the polar and azimuthal angles of $\ns$ in this
system, viz.,
\begin{eqnarray}
{\mathbf n}_s & = & \hat{X}' \sin\psi \sin\Omega +
                   \hat{Y}' \cos\psi -
                   \hat{Z}' \sin\psi \cos\Omega.
\label{eq:ns-orbit-centric}
\end{eqnarray}
Equation~(\ref{eq:ns}) may also be rewritten using the
rotation transformation equations
\begin{eqnarray}
X' & = & X, \\
Y' & = & Y\sin i_o + Z \cos i_o, \\
Z' & = & -Y\cos i_o + Z \sin i_o,
\end{eqnarray}
giving
\begin{eqnarray}
{\mathbf n}_s  &=&  \hat{X}' \sin i_s \sin \lambda  \\ \nonumber
       &&+ \hat{Y}' (\sin i_s \cos \lambda \sin i_o + \cos i_s \cos i_o) + \\\nonumber
       &&+ \hat{Z}' (\cos i_s \sin i_o - \sin i_s \cos\lambda \cos i_o).
\label{eq:ns-observer-centric}
\end{eqnarray}

Setting the components of equation~(\ref{eq:ns-orbit-centric}) equal to
those of equation~(\ref{eq:ns-observer-centric}) we obtain three relations
\begin{eqnarray}
\sin i_s \sin\lambda & = & \sin\psi \sin\Omega \label{eq:angles1} \\
\cos\psi & = & \sin i_s \cos\lambda \sin i_o + \cos i_s \cos i_o
   \label{eq:angles2} \\
\sin\psi \cos\Omega & = & \sin i_s \cos\lambda \cos i_o - \cos i_s \sin i_o
   \label{eq:angles3},
\end{eqnarray}
which will be used in the following sections to derive constraints on
$\psi$ based on measurements of $i_o$ and $\lambda$ and on reasonable
assumptions regarding $i_s$ and $\Omega$.

\section{Given $\psi$, what will RM observations show?}
\label{sec:probind}

Suppose an observer has measured the orbital inclination of a
planetary system to be $i_o$ and is about to measure the RM effect.
If the spin-orbit angle of the system is $\psi$, then what is the
probability distribution for the projected spin-orbit angle $\lambda$
that the observer will measure? In this section we calculate this
function, $p(\lambda | \psi, i_o)$, which will play an important role
in the calculations to follow.

We assume that for a given $\psi$, the probability distribution of the azimuthal angle $\Omega$ is
uniformly distributed between $-\pi$ and $+\pi$. This is self-evident
for a circular orbit, as there is no physical reason to distinguish
any particular azimuth. For an eccentric orbit it is conceivable that
$\Omega$ is correlated with the direction of pericenter, but this
possibility seems unlikely for hot Jupiters, because the torque
exerted on the stellar rotational bulge by the planetary orbit will
cause $\Omega$ to precess. The precession period for a hot Jupiter
orbit is much shorter than the age of the system, and the secular
evolution of both the argument of pericenter and the longitude of the
ascending node---both defined with respect to the stellar equator---is
linear in time (\citealt{2005R}; \S 11.4.2, to lowest order in stellar
shape parameters).  We therefore expect an ensemble of stellar spins
to have a uniform distribution in $\Omega$, even if their orbits are
eccentric. Conversely, the spin-orbit angle $\psi$ remains constant
over the precession cycle (\citealt{2005R}; \S 11.4.2).

To derive $p(\lambda | \psi, i_o)$, we first express $\lambda$ in
terms of $\psi$, $i_o$, and $\Omega$ by eliminating $i_s$ from
equations~(\ref{eq:angles1})-(\ref{eq:angles3}):
\begin{equation}
\lambda(\psi, i_o, \Omega) =
\arctan \left( \frac{\sin\psi \sin\Omega}{\cos\psi\sin i_o + \sin\psi \cos\Omega \cos i_o} \right).
\label{eq:tan-lambda}
\end{equation}
Since $\lambda(-\Omega, \psi, i_o) = -\lambda(\Omega, \psi, i_o)$, to
calculate probabilities of $\lambda$ and $\psi$ we need only consider
$\Omega$ and $\lambda$ in the range $[0,\pi]$.  The results will apply
to negative values of $\lambda$ as positive and negative values occur
with equal probability.

Next, making use of $p(\Omega) = \pi^{-1}$, we transform variables
from $\Omega$ to $\lambda$:
\begin{equation}
\label{eq:transform-to-omega}
p(\lambda | \psi, i_o) = \sum_{i=1}^{N} p(\Omega_i | \psi, i_o) \left| \frac{d\Omega}{d\lambda} \right|_{\Omega=\Omega_i}
                       = \frac{1}{\pi} \sum_{i=1}^{N} \left| \frac{d\Omega}{d\lambda} \right|_{\Omega=\Omega_i}
\end{equation}
where the sum ranges over the $N$ solutions $\Omega_i$ of
equation~(\ref{eq:tan-lambda}), for a given choice of $\lambda$, $\psi$,
and $i_o$.  We now find those solutions.  

\clearpage

\begin{figure*}[htp]
\epsscale{0.8}
\plotone{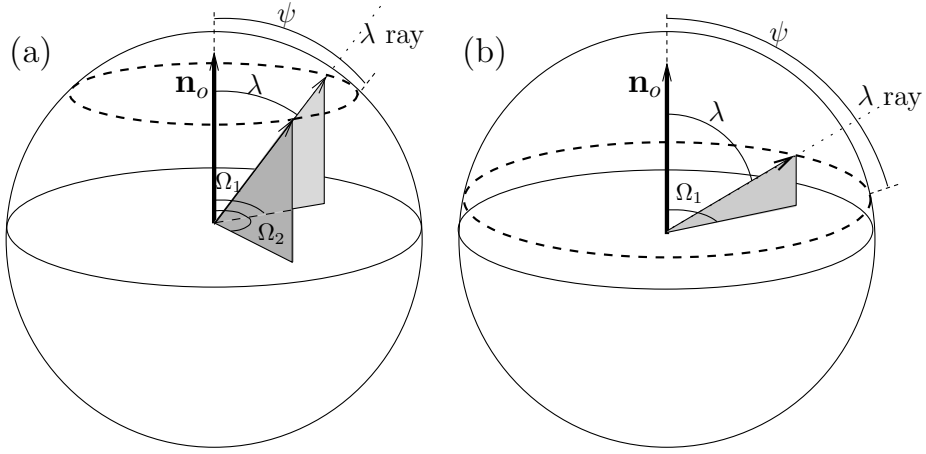}
\caption{Geometry of the roots of equation~(\ref{eq:omegaroots}). For
  given values of $\psi$ and $i_o$ there are either 0, 1, or 2
  possible values of $\lambda$. In these panels the dashed line is the
  ellipse on the sky (the ``$\ns$ ellipse'') that is traced out as $\ns$ (thin vectors) sweeps 
  around $\no$ (thick vector), with $\Omega$ taking on all values, and the dotted line is the ray
  corresponding to a given value of $\lambda$ (the ``$\lambda$ ray'').
  (a) $\psi < i_o$ and $\sin \lambda < \sin\psi / \sin i_o$.  There
  are two intersections of the $\ns$ ellipse and $\lambda$ ray.  (b)
  $i_o < \psi < \pi-i_o$. There is only one intersection.  For $\psi <
  i_o$ and $\sin\lambda > \sin\psi / \sin i_o$ (not shown), the $\ns$
  ellipse is too small to intersect the $\lambda$ ray.}
\label{fig:cone}
\vspace{0.3 in}
\end{figure*}

\begin{figure*}[hp]
\epsscale{1.2}
\plotone{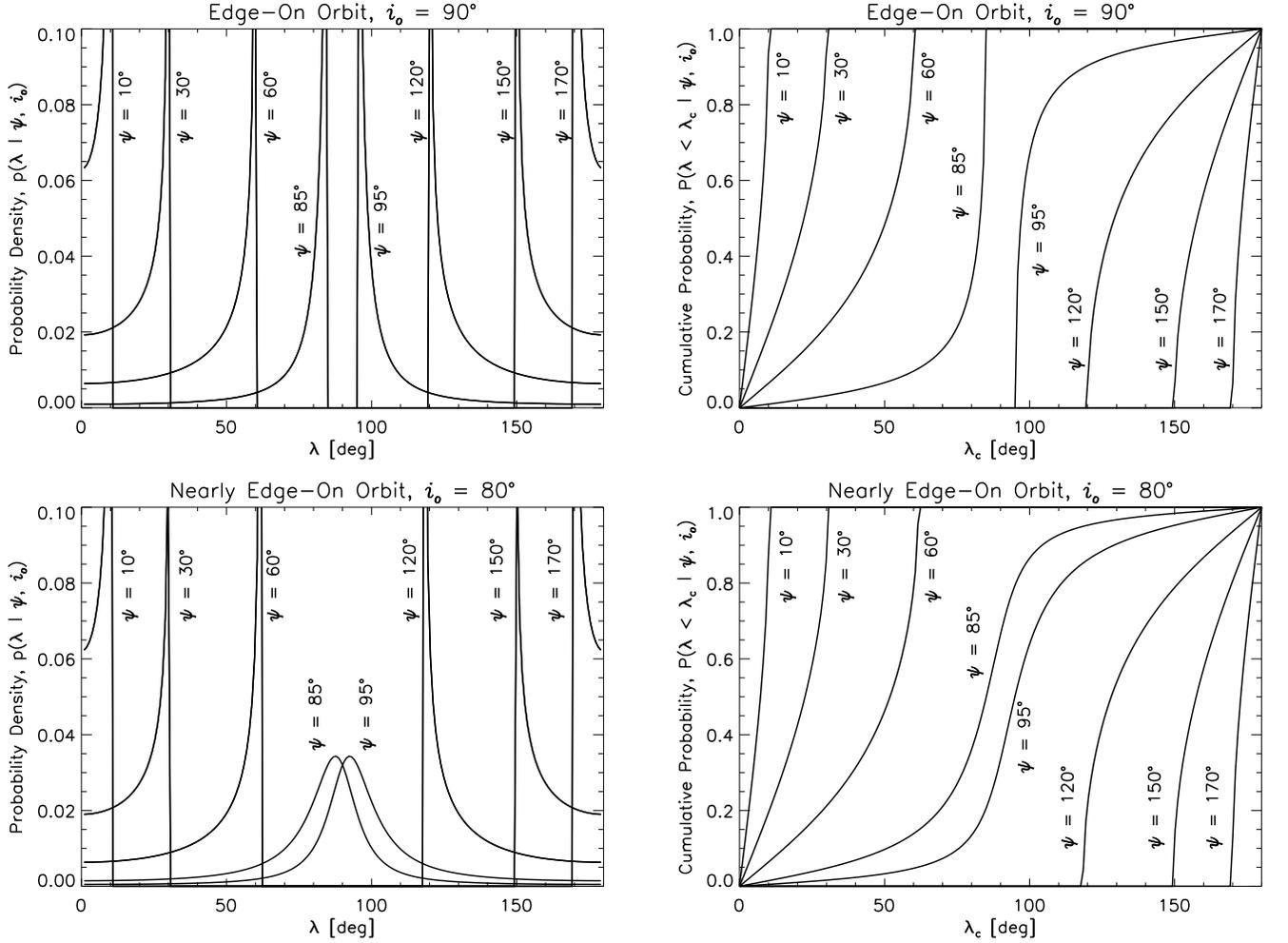}
\caption{Probability distributions for the projected spin-orbit angle $\lambda$
conditioned on the true spin-orbit angle $\psi$ and the orbital
inclination $i_o$. See
Eqs.~(\ref{eq:omegaroots}-\ref{eqn:probdensity-lambda-given-psi-edgeon}).
The upper panels show results for an edge-on orbit ($i_o=90\arcdeg$)
and various choices of $\psi$.  The lower panels show results for a
nearly edge-on orbit ($i_o=80\arcdeg$) and various choices of $\psi$.
\label{fig:p_lambda_given_psi}
}
\vspace{0.3 in}
\end{figure*}

\begin{minipage}[t]{7.0 in}

Letting $u\equiv
\cos\Omega$, equation~(\ref{eq:tan-lambda}) is equivalent to a quadratic
equation in $u$:
\begin{equation}
u^2 \left[ \sin^2\psi(1 + \tan^2\lambda \cos^2 i_o) \right] +
2u  \left[ \tan^2\lambda \sin\psi \cos\psi \sin i_o \cos i_o \right] -
\sin^2\psi \cos^2 i_o .
\end{equation}
When $\psi < i_o$ and $\sin \lambda < \sin\psi / \sin i_o$, this
equation has two roots:
\begin{equation}
u_{1,2} = \cos \Omega_{1,2} =
\frac{
-\tan^2\lambda \cos\psi \sin i_o \cos i_o \pm \sec \lambda \sqrt{\sin^2\psi  - \sin^2 i_o \sin^2\lambda }
}
{
\sin\psi (1 + \tan^2 \lambda \cos^2 i_o)
} . \label{eq:omegaroots}
\end{equation}
Fig.~2a shows the geometrical interpretation of this two-root case.
The ellipse on the sky that is traced out by $\ns$ as $\Omega$ takes
on all values (the ``$\ns$ ellipse'') has two intersections with the
ray that corresponds to the given value of $\lambda$ (the ``$\lambda$
ray''). When $\psi < i_o$ and $\sin\lambda > \sin\psi / \sin i_o$,
there are no real roots of equation~(\ref{eq:omegaroots}). Geometrically,
the $\ns$ ellipse is too small to intersect with the $\lambda$
ray. The last case is $i_o \leq \psi \leq \pi-i_o$, for which
equation~(\ref{eq:omegaroots}) has one real root. Fig.~2b shows an example
of this case. The $\ns$ ellipse encloses the origin, and there is one
intersection point with each $\lambda$ ray. This single root
corresponds to $\Omega_1$ in equation~(\ref{eq:omegaroots}). Thus in
equation~(\ref{eq:transform-to-omega}), $N=0$, $1$, or $2$, depending on
the values of $\psi$, $i_o$, and $\lambda$.

Because of the derivatives $|d\Omega/d\lambda|$ in
equation~(\ref{eq:transform-to-omega}) it is easier to derive the
cumulative distribution $P( \lambda < \lambda_c | \psi, i_o)$ than to
solve directly for $p(\lambda | \psi, i_o)$. The results are:
\begin{equation}
\label{eq:probcum-lambda-given-psi}
P( \lambda < \lambda_c | \psi, i_o) = \left\{
\begin{array}{cl}
1 + \frac{1}{\pi}(\Omega_1 - \Omega_2), &\qquad \psi < i_o~{\rm and}~\lambda_c < \arcsin(\sin\psi/\sin i_o) \\
1, &\qquad \psi < i_o~{\rm and}~\lambda_c \geq \arcsin(\sin\psi/\sin i_o) \\
\frac{\Omega_1}{\pi}, &\qquad i_o \leq \psi \leq \pi-i_o \\
\frac{1}{\pi}(\Omega_1 - \Omega_2), &\qquad \psi > \pi - i_o~{\rm and}~\lambda_c > \pi - \arcsin(\sin\psi/\sin i_o) \\
0, &\qquad \psi > \pi - i_o~{\rm and}~\lambda_c \leq \pi - \arcsin(\sin\psi/\sin i_o)
\end{array}
\right\}
\end{equation}
where $\Omega_{1,2}$ from equation~(\ref{eq:omegaroots}) are evaluated
at $\lambda_c$, $\psi$, $i_o$.

The probability densities are obtained by differentiation. First we
evaluate the derivatives $d\Omega_i/d\lambda$:
\begin{eqnarray}
\frac{d\Omega_{1,2}}{d\lambda} & = & 
\Big( \frac{ 2 \tan\lambda \sec^2\lambda}{\sin \Omega_{1,2} \sin\psi (1 - \tan^2\lambda \cos^2 i_o )} \Big) \nonumber \\
& & \times \Big( \cos\psi \sin i_o \cos i_o + \sin \psi \cos^2 i_o \cos \Omega_{1,2} \pm
\frac{\cos \lambda ( \sin^2 i_o - \sin^2\psi)}{2 ( \sin^2 \psi - \sin^2 i_o \sin^2\lambda)^{1/2}} \Big).
\label{eq:domega-dlambda}
\end{eqnarray}
Finally, we calculate $p(\lambda | \psi, i_o)$ by inserting these
derivatives into equation~(\ref{eq:transform-to-omega}). The full
expressions are too large to reproduce here; instead we give the
expressions into which equation~(\ref{eq:domega-dlambda}) may be
substituted:
\begin{equation}
\label{eq:prob-lambda-given-psi}
p(\lambda | \psi, i_o) = \left\{
\begin{array}{cl}
\frac{1}{\pi} \left(\frac{d\Omega_1}{d\lambda} - \frac{d\Omega_2}{d\lambda} \right), & \sin \psi < \sin i_o~{\rm and}~\sin\lambda < \sin\psi/\sin i_o \\
0, & \sin \psi < \sin i_o~{\rm and}~\sin\lambda \geq \sin\psi/\sin i_o \\
\frac{1}{\pi} \frac{d\Omega_1}{d\lambda}, & \sin \psi \geq \sin i_o 
\end{array}
\right\}
\end{equation}

For a transiting planet, $i_o$ is always close to $\pi/2$. When
$i_o=\pi/2$ exactly, the results are simplified as
\begin{equation}
\label{eqn:probcum-lambda-given-psi-edgeon}
P(\lambda < \lambda_c | \psi, i_o=\pi/2) = \left\{
\begin{array}{cl}
  \frac{2}{\pi} \arccos \left[ \frac{1}{\sin\psi} \left(1-\frac{\cos^2\psi}{\cos^2\lambda_c}\right)^{1/2} \right], &  |\lambda-\pi/2| \geq |\psi-\pi/2| \\
  1, & \psi \leq \lambda_c < \pi/2 \\
  0, & \psi > \lambda_c > \pi/2 \\
\end{array}
\right\}
\end{equation}
and
\begin{equation}
\label{eqn:probdensity-lambda-given-psi-edgeon}
p(\lambda | \psi, i_o=\pi/2) = \left\{
\begin{array}{cl}
  \frac{2}{\pi} \frac{ \cos \psi }{ \cos \lambda (\cos^2 \lambda - \cos^2 \psi)^{1/2} }, & |\lambda-\pi/2| \geq |\psi-\pi/2| \\
  0, & |\lambda-\pi/2| < |\psi-\pi/2|
\end{array}
\right\}
\end{equation}
In the degenerate case $i_o=\pi/2$ and $\psi=\pi/2$, $\lambda$ is
observed to be either $-\pi/2$ or $\pi/2$ with equal probability.

Figure~\ref{fig:p_lambda_given_psi} shows the probability densities and
cumulative distributions for $i_o=90\arcdeg$ and $i_o=80\arcdeg$ and
some representative values of $\psi$. To gain an intuitive
appreciation of the results, consider an edge-on orbit with $\psi <
90\arcdeg$, shown in the left halves of the upper two panels. In this
case the spin-orbit angle $\psi$ is an upper bound on its sky-projected version
$\lambda$. For $\psi = 30\arcdeg$, the chance of observing $\lambda$ to
be smaller than $\psi$ by a factor of two is approximately 35\%. In
contrast, for $\psi = 85\arcdeg$, the chance of observing $\lambda$
smaller than $\psi$ by a factor of 2 is only $\approx$5\%. In this
sense, $\lambda$ is a more faithful indicator of $\psi$ when $\psi$ is
large. For non-edge-on orbits (the lower panels), the maximum value of
$\lambda$ is increased, and for a non-edge-on orbit with $\psi$ near
$90\arcdeg$, it is possible to observe any value of $\lambda$.

\end{minipage}

\clearpage

\subsection{Given $\lambda$ from RM observations, what may be inferred about $\psi$?}
\label{sec:invert}

\begin{figure}[htbp]
\epsscale{1.2}
\plotone{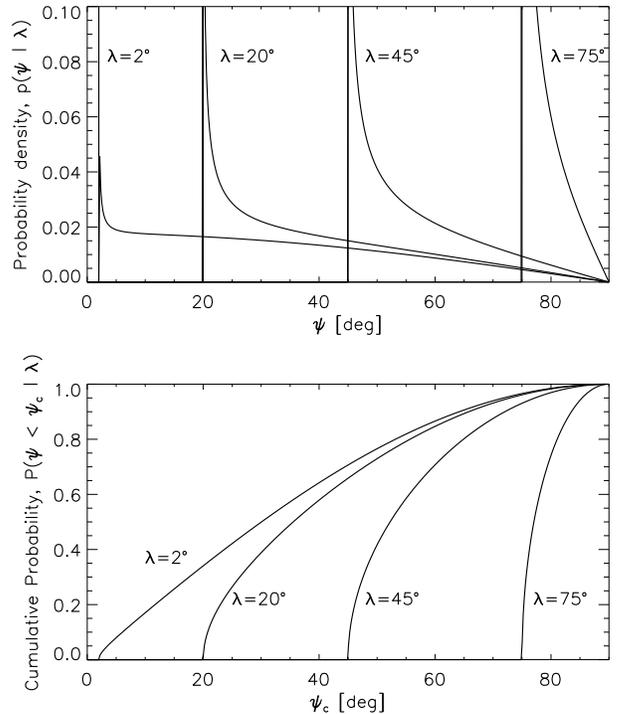}
\caption{ Probability distributions for the true spin-orbit angle
  $\psi$ conditioned on the projected spin-orbit angle $\lambda$,
  assuming an edge-on orbit ($i_o=90\arcdeg$) and random spin-orbit
  alignment. See
  equations~(\ref{eq:prob-psi-given-lambda-edgeon})-(\ref{eq:probcum-psi-given-lambda-edgeon}).
}
\label{fig:p_psi_given_lambda}
\vspace{0.3 in}
\end{figure}

Suppose an observer has just measured $i_o$ and $\lambda$ for a
particular transiting system. What may be reasonably inferred about
the spin-orbit angle $\psi$? We appeal to Bayes' theorem:
\begin{equation}
p(\psi | \lambda, i_o) \propto p( \lambda | \psi, i_o ) p( \psi ),
\end{equation}
where $p(\lambda | \psi, i_o)$ was calculated in the previous section,
and $p(\psi)$ is the ``prior'' distribution, quantifying the
observer's assumptions prior to the measurement.  In this section we
adopt a prior distribution $p(\psi) = \sin\psi$, implying that $\ns$
and $\no$ are uncorrelated and $\ns$ is randomly oriented in space.
This is the most uninformative or conservative assumption, in the
sense that if $\no$ and $\ns$ are instead highly correlated (with
consequent implications for the theory of planet migration or tidal
evolution), this fact should be demonstrated based on the data, rather
than assumed from the outset. Hence, $p(\psi | \lambda, i_o)$ may be
obtained by multiplying equation~(\ref{eq:prob-lambda-given-psi}) by $\sin
\psi$ and renormalizing. For brevity, we give here only the analytic
results for the case of an edge-on orbit and $\lambda < \pi/2$,
measured with no error:
\begin{equation}
\label{eq:prob-psi-given-lambda-edgeon}
p(\psi | \lambda, i_o=\pi/2) = \left\{
\begin{array}{cl}
0, & \psi < \lambda \\
\frac{ \cos \psi \sin \psi }{ \cos \lambda (\cos^2 \lambda - \cos^2 \psi)^{1/2}}, & \psi \geq \lambda
\end{array}
\right\}
\end{equation}
and the corresponding cumulative probability function is
\begin{equation}
\label{eq:probcum-psi-given-lambda-edgeon}
P( \psi < \psi_c | \lambda, i_o=\pi/2) = \left\{
\begin{array}{cl}
0, & \psi_c < \lambda \\
\Big(1 - \frac{ \cos^2 \psi_c }{ \cos^2 \lambda } \Big)^{1/2}, & \psi_c \geq |\lambda|
\end{array}
\right\}
\end{equation}
These results are plotted in Figure~\ref{fig:p_psi_given_lambda} for
some representative choices of $\lambda$.  When $\lambda$ is observed
to be small, the {\it a posteriori}\, probability distribution of
$\psi$ has a very narrow spike near $\lambda$ and extends broadly from
$\psi = \lambda$ to $90\arcdeg$. When $\lambda$ is observed to be
large, it is more likely that $\psi$ is close to $\lambda$.  Just as
$M_p$ cannot be constrained strongly by $M_p\sin i_o$ for a
Doppler-detected planet, we find that $\psi$ cannot be constrained
strongly by $\lambda$ for an RM-detected planet, although the nature
of the constraint is more complex in the latter case. In particular, for the edge-on case,
it is possible to distinguish whether an orbit prograde or retrograde
without ambiguity, even though the value of $\psi$ is quite uncertain.

As mentioned in \S~2, in principle one may obtain the missing
information about $i_s$ via asteroseismology, or via the combination
of estimates of the stellar radius $R_s$, projected rotation velocity
$v\sin i_s$, and stellar rotation period $P_s$. Although
asteroseismology has never been undertaken for a transiting
exoplanetary system, the other method has been employed for HD~189733
\citep{2007Wa, 2008HW} and CoRoT-Exo-2 \citep{2008Bo}, and in both
cases it was found that $\sin i_s$ is consistent with $1$ (i.e., the
equator is edge-on). However near $\sin i_s=1$ a small error in the
measured $\sin i_s$ leads to a big error in $i_s$. Therefore the data
exclude only highly misaligned systems ($\psi \gsim 45\arcdeg$). In
the calculations to follow regarding the entire ensemble, we chose not
to make use of these constraints specific to HD~189733 and
Corot-Exo-2, for simplicity and because the extra information does not
lead to significantly more powerful constraints.

Even when the rotation period is not available, one may exclude very
nearly pole-on configurations of the star because they would require
the star to be rotating unrealistically rapidly. Certainly the
rotation rate cannot exceed the breakup speed, although in practice we
find that in realistic cases, applying this constraint does not make a
perceptible difference in the distribution for $\psi$.  One might go
further by applying an {\it a priori}\, constraint on $i_s$ to enforce
agreement with the ``typical'' rotation rate for a star of the given
spectral type and age. For the present study we chose not to apply any
such constraint, to avoid complications due to the uncertainties in
the stellar types and ages, the intrinsic scatter in rotation rates,
and the possibility that the rotation rates of stars hosting close-in
giant planets may be systematically different from stars in general
(due to tidal torques, earlier generations of ``swallowed'' planets,
or other unforeseen effects).

\section{Inferences from an ensemble of systems}
\label{sec:ensemble}

We have seen that $\psi$ cannot be tightly constrained in an
individual system, even when $\lambda$ has been measured to within a
few degrees, and even when $\sin i_s$ is constrained by a measured
rotation period, stellar radius, and projected rotation rate. The
purpose of this section is to derive stronger constraints by combining
the results from different systems. The first such attempt, by
\cite{2006W}, demonstrated that the 3 measurements of $\lambda$
available at that time were strongly inconsistent with an isotropic
distribution of spin-orbit angles. This conclusion has been
strengthened with the addition of many more systems with small values
of $|\lambda|$, and it is now clear that $\ns$ and $\no$ are
correlated. The next natural question is: given the RM data, what is
the distribution of spin-orbit angles? For instance, (a) is there a
``typical'' value which describes the mean and dispersion, and (b) is
a single smooth distribution a good description of the data, or is
there evidence for more than one population?

To answer these question, we suppose that the spin-orbit angles of the
systems under consideration were drawn from a probability distribution
$p(\psi)$ (the ``model''), and we use the data to constrain the
mathematical form of $p(\psi)$. \cite{2006W} already showed that
the isotropic model, $p_{\rm I} = \onehalf \sin\psi$, is untenable. A
good theory of planet formation, migration, and evolution should be
able to supply $p(\psi)$, or at least its general form. We will not
attempt to develop such a theory here. Instead we will use simple
mathematical forms of $p(\psi | \mathbf{a})$ with a few free
parameters and derive the probability distribution for those
parameters, conditioned on the data.

Let the model parameters form a vector $\mathbf{a}$. The data consist
of measurements of $\lambda$ and $i_o$ for $N_s=11$ systems. These
``measurements'' are themselves probability distributions for
$\lambda$ and $i_o$. We neglect the error in $i_o$, and denote by
$p_{{\rm obs},k}(\lambda)$ the probability distribution for $\lambda$
based on the observations of system $k$ (from Table~\ref{table:rm}).
We approximate the measurements as Gaussian distributions with the
quoted $\lambda$ and $\sigma_{\lambda}$ as the mean and standard
deviation.\footnote{In some cases, even when the radial-velocty
  measurement errors are Gaussian, the posterior distribution $p_{{\rm
      obs},k}(\lambda)$ is not Gaussian. This is especially true of
  systems with slow stellar rotation rates or small transit impact
  parameters [see, e.g., the TrES-2 system \cite{2008Wa} or the
  HAT-P-2 system \cite{2007Wb,2008L}]. We investigated the sensitivity
  of our results on the assumption of a Gaussian distribution in
  $\lambda$ by using the actual posterior distribution for $\lambda$
  whenever we had enough information to compute it.  We found that the
  ensemble results were not significantly affected, because the most
  non-Gaussian cases were those with large errors, which were already
  downweighted in the Bayesian analysis.} The model
$p(\psi|\mathbf{a})$ implies a certain probability distribution for
$\lambda$, given by
\begin{equation}
p'(\lambda | i_o,\mathbf{a}) = \int_0^\pi~p(\lambda | \psi, i_o) p(\psi|\mathbf{a})~d\psi, \label{eq:pfisher}
\end{equation}
where $p(\lambda | \psi, i_o)$ is given by
equation~(\ref{eq:prob-lambda-given-psi}). In practice this integral
is problematic because it integrates over the singularities visible in
Figure~\ref{fig:p_lambda_given_psi}, but we found that the singularity
handlers in \emph{Mathematica} are able to perform the integral
numerically. For edge-on orbits ($i_o=\pi/2$), the simplified version
of $p(\lambda | \psi, i_o=\pi/2)$ given by
equation~(\ref{eq:prob-psi-given-lambda-edgeon}) is applicable, and
the transformation
\begin{equation}
y = \Big( 1 - \frac{ \cos^2\psi }{ \cos^2 \lambda } \Big)^{1/2}
\label{eq:nonsingular-transformation}
\end{equation}
removes the singularity. This provided a useful check on the ability
of \emph{Mathematica} to handle the singularities; for $i_o=\pi/2$ the
numerical integrals were identical whether or not the transformation
of equation~(\ref{eq:nonsingular-transformation}) was employed.

We may write the conditional probability as
\begin{equation}
p({\rm data} | \mathbf{a}) = \prod_{k=1}^{N_s} \int_{-\pi}^{+\pi} p_{{\rm obs},k}(\lambda) p'(\lambda | i_{p,k}, \mathbf{a}) d\lambda,  \label{eq:pprod}
\end{equation}
and then use Bayes' theorem,
\begin{equation}
p(\mathbf{a} | {\rm data}) \propto p( {\rm data} | \mathbf{a} ) p( \mathbf{a} ), \label{eq:ourbayes}
\end{equation}
where $p(\mathbf{a})$ is the prior probability density that is
assigned to the parameters $\mathbf{a}$. Next, let us choose
distributions to test.

\subsection{A Fisher distribution}
\label{subsec:fisher}

\begin{figure}[t]
\epsscale{1.0}
\plotone{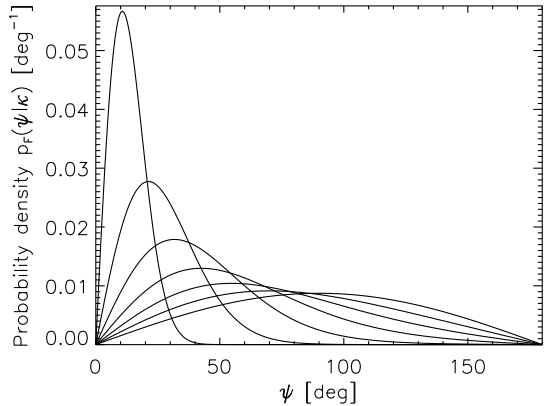}
\caption{ 
The Fisher probability distribution (see eq.~[\ref{eq:fisher}]), for
some representative values of the concentration parameter $\kappa$.
For $\kappa\rightarrow 0$ the Fisher distribution becomes an isotropic
distribution ($\onehalf \sin \psi$), and for $\kappa\rightarrow\infty$
and $\psi \rightarrow 0$ the Fisher distribution becomes a Rayleigh
distribution with width parameter $\sigma = \kappa^{-1/2}$. The
displayed curves have $\kappa = 0.0$, $0.4$, $0.9$, $1.6$, $3.1$,
$7.1$, and $29$. These values were chosen because our prior,
$p(\kappa) \propto (1+\kappa^2)^{-3/4}$, assumes equal probability for
each interval between adjacent values of $\kappa$.
\label{fig:fisher}
}
\vspace{0.3 in}
\end{figure}

If $\psi$ were a Cartesian coordinate instead of a polar angle, one
might model $p(\psi)$ as a Gaussian distribution with zero mean and
variance $\sigma$, and derive the probability distribution for
$\sigma$ conditioned on the data. In the theory of directional
statistics, the function that plays the same widespread and suitably
generic role as the Gaussian distribution is the \cite{1953F}
distribution,
\begin{equation}
\label{eq:fisher}
p_{\rm F}(\psi | \kappa) = \frac{\kappa}{2 \sinh \kappa} \exp ( \kappa \cos \psi ) \sin \psi, \label{eq:fisherdist}
\end{equation}
where $\kappa$ is the concentration parameter, a measure of the
concentration of the probability distribution around $\psi=0$. For
$\kappa = 0$, the distribution becomes the isotropic distribution
$p_{\rm I}(\psi) = \onehalf \sin \psi$.  For $\kappa \gg 1$ and $\psi
\rightarrow 0$, the distribution becomes a Rayleigh distribution
\begin{equation}
p_{\rm R}(\psi|\sigma) = \frac{\psi}{\sigma^2} \exp \left( - \frac{\psi^2}{2 \sigma^2} \right) \label{eq:rayleighdist}
\end{equation}
with a width parameter $\sigma = \kappa^{-1/2}$.  Mathematical
properties of the Fisher distribution, and its relation to other
distributions, can be found in \cite{1982W}.  Some examples of Fisher
distributions are plotted in Figure~\ref{fig:fisher}.

Let us assume that the spin-orbit angles are drawn from a Fisher
distribution, and derive the probability distribution for $\kappa$
conditioned on the data.  We choose a prior distribution $p(\kappa)
\propto (1+\kappa^2)^{-3/4}$. This has the desirable limits $p(\kappa)
\rightarrow \rm{constant}$ for $\kappa \rightarrow 0$ (for broad
distributions it is uninformative in $\kappa$), and $p(\sigma)
\rightarrow \rm{constant}$ for $\kappa \rightarrow \infty$ (for
narrow, Rayleigh-like distributions it is uninformative in $\sigma =
\kappa^{-1/2}$). The particular values of $\kappa$ for which the
distributions are illustrated in Figure~\ref{fig:fisher} were chosen
because each interval in $\kappa$ is equally likely, according to our
prior.

\begin{figure}[t]
\epsscale{1.1}
\plotone{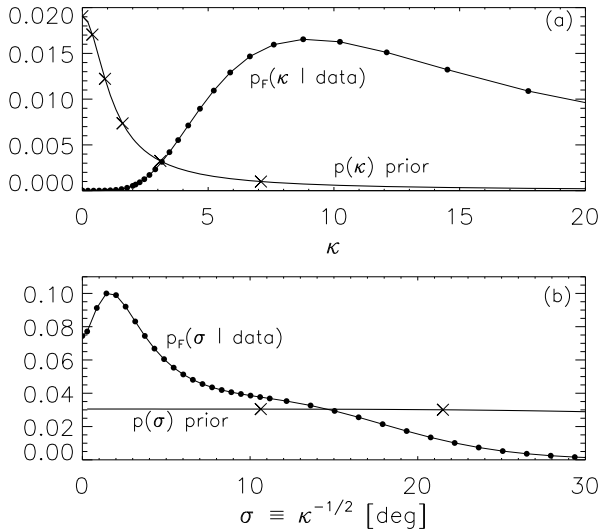}
\caption{
Modeling the RM ensemble with a Fisher distribution.
(a) The line with dots shows the probability of the concentration
parameter $\kappa$, conditioned on the data.  The dots show the
specific values of $\kappa$ for which we computed the posterior
probability.  The line with \emph{x}'s shows the assumed prior
distribution of $\kappa$.  The \emph{x}'s show the values of $\kappa$
for which the Fisher distributions are illustrated in
Figure~\ref{fig:fisher}.
(b) Correspondence with the more familiar Rayleigh distribution
(eq.~[\ref{eq:rayleighdist}]), using the equivalence $\sigma \equiv
\kappa^{-1/2}$ that is motivated by the high-$\kappa$, low-$\sigma$
limit. Lines and points have the same meaning as in panel (a).  In
neither panel are the prior distributions normalized. The posterior
distributions are normalized by the ``evidence'' for the Fisher
distribution model (see equation~\ref{eq:evidence}).
\label{fig:resfisher}
}
\vspace{0.3 in}
\end{figure}

The probability density for each value of $\kappa$, conditioned on the
data, is the product of the prior and the probability density of the
data given the value of $\kappa$, according to
equation~(\ref{eq:ourbayes}).  Figure~\ref{fig:resfisher}(a) shows the
resulting function $p(\kappa| \rm{data})$, based on the 11 available
RM measurements. It has been suitably normalized to unit
probability. [The prior $p( \kappa )$ is also displayed in
Fig.~\ref{fig:resfisher}(a), for reference.]  Based on this result, we
find that the characteristic concentration parameter is $\kappa > 7.6$
with 95\% confidence.

The results for $\kappa$ can be converted to a characteristic angular
dispersion $\sigma$ (in degrees) using $\sigma \equiv \kappa^{-1/2}$,
bearing in mind that the Fisher distribution with concentration
parameter $\kappa$ is the same as the Rayleigh distribution with width
parameter $\kappa^{-1/2}$ in the limit of $\kappa \rightarrow \infty$.
The resulting distributions $p_{\rm F}( \sigma | \rm{data} )$ and $p(
\sigma )$ are plotted in Figure~\ref{fig:resfisher}(b). The width
parameter $\sigma$ is less than $22\arcdeg$ with $95\%$ confidence.

We now examine the sensitivity of these results to certain aspects of
the input data. First, we recompute the results without including the
XO-3 data. This is because the finding of a strong misalignment in
that system was considered tentative by the observers themselves. A
standard RM model does not provide a statistically acceptable fit to
the XO-3 data \citep{2008H}. It may be relevant that some of the data
were contaminated by bright moonlight, requiring significant
corrections to be applied to the spectra, and some of the data were
taken at very high air masses.

\begin{figure}[t]
\epsscale{1.0}
\plotone{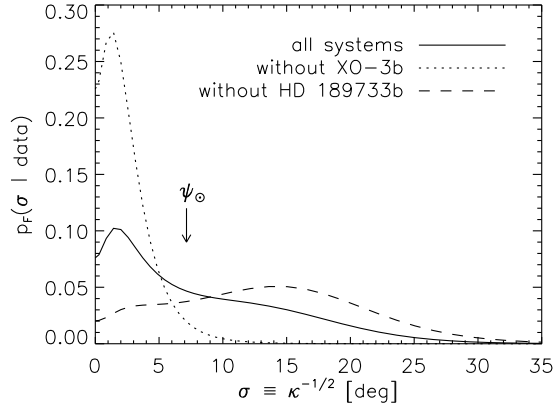}
\caption{ The same as Figure~\ref{fig:resfisher}(b), but restricting the
analysis to 10 systems instead of 11, by removing either XO-3 or
HD~189733. The results are very different in those cases.  Thus the
results depend strongly on the few systems for which $\lambda$ is
measured with the best precision, and the single system with an
apparently large misalignment.  The Solar value of $\psi$ is shown for
reference.
\label{fig:respriored_mam}
}
\vspace{0.3 in}
\end{figure}

Figure~\ref{fig:respriored_mam} show the results when XO-3 is ignored
and the other 10 systems are included as before. The results are very
different: the distribution is tightly constrained near zero: $\sigma
< 7^\circ$ with $95\%$ confidence. That these ten systems are
consistent with perfect alignment ($\sigma = 0$) can be seen by
computing $\chi^2 = \sum (\lambda/\sigma_\lambda)^2$ from
Table~\ref{table:rm}: it is $11.7$, with 10 degrees of freedom.
Therefore, apart from XO-3, the non-zero values of $\lambda$ are
consistent with observational errors alone.  The $p(\sigma)$
distribution maximum likelihood is at non-zero $\sigma$ because the
reduced $\chi^2$ is greater than 1, but this departure from $\sigma=0$
is not statistically significant.

Second, we investigate how sensitive are the results to the few most
precise RM measurements. The dashed line in
Fig.~\ref{fig:respriored_mam} show the results if we ignore HD~189733,
the system with the smallest error bar $\sigma_\lambda$, and include
all 10 of the remaining systems.  In this case, the XO-3 result has
enough statistical weight to pull the peak of the distribution
strongly away from zero. Also plotted for reference is $\psi_\odot$,
the Solar spin-orbit angle. It must be remembered, though, that
$\psi_\odot$ is a particular spin-orbit angle while $\sigma$ describes
the dispersion of $\psi$. The Solar value of $\psi_\odot$ is typical
of the ``allowed'' spin-orbit angles of hot Jupiters with their host
stars.

With an eye towards future statistical analyses using the Fisher
distribution, we note that the most computationally challenging aspect
of the analysis was performing the integral~(\ref{eq:pfisher}). A
major simplification is available for nearly edge-on orbits, if the
data are already known to favor a highly concentrated distribution
($\kappa \gg 1$). In this case the hypothesized distribution for
$\psi$ is a Rayleigh distribution of parameter $\sigma \ll 1$, and the
distribution of $\lambda$ is Gaussian with a standard deviation
$\sigma$. The problem of constraining the distribution of $\psi$ is
reduced to the problem of determining the true standard deviation of a
distribution from which several noisy data points have been drawn.
The dispersion $\sigma$ can be estimated by finding the value of
$\sigma$ that gives $\chi^2=N_s$ when it is added in quadrature with the
measurement errors. The ten systems besides XO-3 are in this simple
regime. The simplified procedure described in this paragraph gives
$\sigma=1.1^\circ$, in agreement with the maximum likelihood value
given for those ten systems in Figure (\ref{fig:respriored_mam}).

\subsection{A sum of two distributions:  isotropic and perfectly-aligned}
\label{subsec:twodist}

An alternative and equally simple way to describe the data is to
suppose that all systems are drawn either from an isotropic
distribution (with probability $f$) or from a very well-aligned
distribution (with probability $1-f$). We further suppose that the
well-aligned distribution is sufficiently concentrated around $\psi=0$
that none of the current measurements would be able to distinguish it
from a delta function. This toy model will be a useful baseline for
limiting the fraction of planets that migrate by various channels,
some of which yield a nearly isotropic distribution of spin-orbit
angles (specific examples are cited in \S\ref{sec:discussion}).

\begin{figure}[t]
\epsscale{1.0}
\plotone{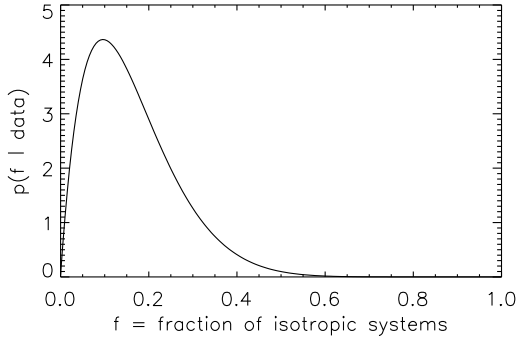}
\caption{Modeling the RM ensemble with the sum of an isotropic
  distribution and a perfectly-aligned (delta-function) distribution.
  A fraction $f$ of the systems are drawn from the isotropic
  distribution, and the remaining fraction $(1-f)$ are drawn from the
  delta-function distribution.  Plotted is the posterior probability
  distribution for $f$, given the data.  We find $f<0.36$ with
  $95\%$ confidence.}
\label{fig:fdist}
\vspace{0.3 in}
\end{figure}

The probability of the data ($\lambda_k$, $\sigma_{\lambda,k}$) given
such a model is:
\begin{equation}
p({\rm data} | f) = 
\prod_{k=1}^N \Big[
f \frac{1}{2\pi} + (1-f) \frac{2}{\sqrt{2\pi\sigma_k^2}}
\exp \left( -\frac{\lambda_k^2}{2 \sigma_{\lambda,k}^2} \right)
\Big].
\end{equation}
The factor of $2$ in the numerator of the Gaussian portion above
arises from the $\pm\lambda$ degeneracy mentioned in \S\ref{sec:geometry}.
Both of the terms in the sum are independent of $i_o$. Adopting a
uniform prior for $f$, and using Bayes' theorem, we plot the result
for $p(f | {\rm data})$ in Figure~\ref{fig:fdist}. The data demand
that $f<0.36$ with $95\%$ confidence. The favored value of $f$ is 0.1,
implying that approximately one system out of 11 is drawn from
the isotropic distribution, clear indication that this result is being 
driven by XO-3.  If we remove XO-3 from the analysis (to check
the sensitivity of the analysis to this one system), $f<0.25$ at 
$95\%$ confidence, and the maximum likelihood is $f=0$ exactly.

\subsection{A sum of two Fisher distributions}
\label{subsec:twofish}

Another possible model is a sum of two Fisher distributions with
different concentration parameters. This could describe, for instance,
two different channels by which giant planets migrate to close-in
orbits, which produce different final distributions of spin-orbit
angles. \cite{2001B} has found that a sum of two Fisher-like
distributions is needed to fit the inclination distribution of Kuiper
belt objects, and this multi-component model has been a very useful
constraint on dynamical theories. In this ``two-Fisher'' model, a
fraction $f$ of systems are Fisher-distributed with concentration
parameter $\kappa_1$, and the remaining fraction $(1-f)$ of systems
are Fisher-distributed with concentration parameter $\kappa_2$. Thus
the two-Fisher model has three free parameters. The results of the
previous section correspond to the case $f \approx 0.1$, $\kappa_1=0$,
$\kappa_2\rightarrow\infty$.

To determine the joint posterior probability distribution for $f$,
$\kappa_1$, and $\kappa_2$, we computed
\begin{eqnarray}  
p_{2{\rm F}}(f, \kappa_1, \kappa_2 | {\rm data} ) &=& p(f) p(\kappa_1) p(\kappa_2) \times  \\ \nonumber
& &\Big[ f p_F({\rm data} | \kappa_1) + (1-f) p_F({\rm data} | \kappa_2) \Big] .
\end{eqnarray}
The difficult integrals implied by this equation were already computed
for the single-Fisher model. The results are $f=0.10^{+0.13}_{-0.05}$,
$\kappa_1=0.22^{+1.04}_{-0.22}$, and $\kappa_2=110^{+230}_{-76}$. The
first Fisher distribution is consistent with an isotropic
distribution. The second Fisher distribution is approximately
equivalent to a Rayleigh distribution with width parameter
$\sigma_2=5.5 ^{+4.3}_{-2.4}$~deg. The probability contours show
that the correlations between the parameters are relatively small, 
especially within the 1-$\sigma$ preferred region.

\subsection{Choosing among different models} \label{sec:modelchoose}

The results of the 3-parameter, two-Fisher model are consistent with
the results of the single-parameter, isotropic~$+$~perfectly-aligned
model given in the previous section. The greater complexity of the
two-Fisher model makes it less appealing. This loss of appeal can be
quantified within a Bayesian framework, which has a quantitative
expression of Occam's razor.\footnote{A lucid discussion of Bayesian
  model choosing is given by \citealt{2003MacKay}, ch.~28.}  Each
model has an associated ``evidence,''
\begin{equation}
E \equiv \int p({\rm data}  | \mathbf{a}) p(\mathbf{a}) d\mathbf{a}, \label{eq:evidence}
\end{equation}
where $\mathbf{a}$ is the vector of model parameters. This is the
normalization factor used in Bayes' theorem, which turns
proportionality~(\ref{eq:ourbayes}) into an equation. The presence of
the first factor, $p({\rm data} | \mathbf{a})$, is the quantitative
expression of the intuition that the data may be taken as evidence for
the model only when the model predicts the data are probable. The
second factor, $p(\mathbf{a})$, assigns greater evidence to models
that concentrate their predictive power in the region where the data
are found. This is because the integral over the prior distribution
$p(\mathbf{a})$ is normalized to unity; if a prior distribution is
spread too thinly over the parameter space of $\mathbf{a}$, then it
cannot give much weight to models that are consistent with the data.

We computed the evidence for the models described in the three
previous sections. The single-Fisher model has $E=14.4$, the
isotropic~$+$~perfectly-aligned model has $E=1927$, and the two-Fisher
model has $E=726$.  The model that mixes the two extreme distributions
does the best. It is favored by a factor of $\sim 130$ over the
single-Fisher model.

The difficulty with the single-Fisher model is that it cannot
simultaneously account for the majority of systems that are
well-aligned while also including XO-3. A small value of $\kappa$
makes XO-3 probable, but the 10 other systems are somewhat less
probable, and the multiplication of these 10 lessened probabilities
according to equation~(\ref{eq:pprod}) means there is little evidence
for small $\kappa$. A large value of $\kappa$ makes most of the data
probable, but then the XO-3 result is very improbable, and the result
is poor evidence for large $\kappa$. The distribution that mixes an
isotropic distribution and a perfectly-aligned distribution overcomes
this difficulty by allowing \emph{both} the majority of systems \emph{and} XO-3 to
be reasonably probable. The only free parameter is $f$, and the
constraints on $f$ are loose.  Thus, the data provide substantial
evidence for a large fraction of the parameter space.

The poor showing of the two-Fisher model relative to the
isotropic~$+$~perfectly-aligned model indicates that the two-Fisher
model is a needless complication.  The extra two parameters open up
two more dimensions in the model's parameter space. As a result, much
of the prior probability is ``wasted'' on regions of parameter space
that are ruled out by the data. We hope that some day there will be
enough high-quality RM data to justify a more complicated model such
as the two-Fisher model. In this regard we note that 379 Kuiper belt
objects were discovered before their inclination distribution was
modeled this way \citep{2001B}.

We have already shown that the XO-3 measurement has an especially
strong influence on the results. Unfortunately, as mentoned earlier,
this measurement is considered suspect because of the possibility of
systematic errors.  Following the referee's suggestion, we may bring
this suspicion under the umbrella of the Bayesian analysis by
including a ``degree of belief'' parameter $p_{\rm r}$, which gives
the {\it a priori}\, probability that the XO-3 measurement will prove
to be correct. Since the nature of the systematic errors (if any) is
not known, $p_{\rm r}$ is rather subjective and open to debate. Our
goal is not to determine the value of $p_{\rm r}$, but rather to ask
what is the minimum value of $p_{\rm r}$ that is required for our
conclusion to hold that the isotropic~$+$~perfectly-aligned model is
preferred.

Each of the three models---single-Fisher, perfectly-aligned plus an
isotropic fraction, and two-Fisher---is fitted to the 11-member
ensemble including XO-3, and also fitted to the 10-member ensemble
excluding XO-3. Then the evidence for each model is computed as a
weighted sum, with weight $p_{\rm r}$ applied to the 11-member set and
weight $(1-p_{\rm r})$ to the 10-member set. The evidence values of
the models are computed and compared as a function of $p_{\rm r}$. The
result is that for $p_{\rm r} < 0.95$, all three models fit the data
equally well; they have evidence values within a factor of 3 of each
other. We conclude that unless one has $>$95\% confidence that the
XO-3 result is robust, then a single smooth distribution of $\psi$ is
a perfectly viable description of the ensemble.

\section{Discussion}
\label{sec:discussion}

The angle between $\ns$ and $\no$ is a fundamental geometric property
of exoplanetary systems. A good theory of planet formation, migration,
and evolution ought to predict the statistical relationship between
$\ns$ and $\no$ for hot Jupiters. This relationship is potentially
measurable via RM observations. In this paper we have overcome an
inherent limitation of RM observations---they are sensitive only to
the angle between the sky projections of the orbital axis and the
stellar rotation axis---by showing how to analyze the whole ensemble
in a Bayesian framework.

We modeled the 11 published measurements using a Fisher distribution,
and found that the concentration parameter $\kappa > 7.6$ with 95\%
confidence. In this limit of a rather concentrated distribution, the
Fisher distribution is equivalent to a Rayleigh distribution with
width parameter $\sigma=\kappa^{-1/2}$. Based on the 11 data points,
$\sigma < 22\arcdeg$ with 95\% confidence.  For comparison, the Solar
obliquity is $7\arcdeg$.  If we set aside XO-3 (for instance, if the
``tentative'' detection of a strong misalignment is contradicted by
higher-precision data), then the width parameter is $<6.6^\circ$ with
95\% confidence. In that case the hot Jupiters are just as
well-aligned as the Solar system.

The 11 data points also provide statistical evidence for two distinct
populations within the ensemble, which might be interpreted as two
different migration channels. Specifically, a model in which the
systems are drawn from the sum of isotropic and perfectly-aligned
distributions fits the data better than a model with a single smooth
distribution (\S~\ref{sec:modelchoose}). This is a reasonable
conclusion, given that XO-3 shows the only evidence for a strong
misalignment. However, due to the projection effect in converting from
$\psi$ to $\lambda$, it was not obvious prior to our analysis that the
XO-3 result cannot be accomodated as part of the ``tail'' in a smooth
distribution of spin-orbit angles. In fact the data \emph{do not}
imply that XO-3 is the \emph{only} system in the ensemble that is
likely to be drawn from an isotropic distribution. We conclude only
that fewer than 36\% of the systems are drawn from an isotropic
distribution, with 95\% confidence. It is possible that several
members of the ensemble were drawn from an isotropic distribution.

There is plenty of room for improvement in the quantity and quality of
the RM data that are needed for this type of study. We have found that
the present data are sufficient only to constrain single-parameter
models for the ensemble. We also showed that the current results are
highly sensitive to the few systems with the finest measurement
precision. Measurements of $\lambda$ with a precision of a few degrees
are still in high demand. The ``tentative'' result for XO-3 needs to
be followed up with more definitive data, as that single result weighs
heavily on the Bayesian calculation. In addition, a uniform analysis
of the data across many systems would be useful. Among the nonuniform
aspects of the data analyses are whether or not $v\sin i$ was treated
as a free parameter or subject to a constraint based on the observed
line-broadening; whether or not the uncertainties in the photometric
transit parameters were taken into account; whether or not correlated
noise in the radial-velocity data was assessed and taken into account
if necessary; and whether or not the effects of spectral deconvolution
and cross-correlation algorithms were calibrated. (These algorithms
need calibration because they are generally intended to measure
Doppler shifts, rather than model the actual RM spectral distortion
that only superficially resembles a Doppler shift.)

There is also a potential bias regarding which systems are selected
for measurement of the RM effect, as pointed out by \cite{2008Wa}.
Stars with low $v \sin i$ tend to be avoided, as the radial velocity
anomaly would be small and the achievable precision in $\lambda$ is
comparatively poor.  However, stars with low $v\sin i_s$ are more
likely to be viewed pole-on and therefore have a large $\psi$. In the
present work we have not attempted to correct for such a selection
effect.

Despite the limitations of the current data, the relative success of
the two-component model (isotropic~$+$~perfectly-aligned) leads us to
speculate on the implications for theories that attempt to explain the
presence of hot Jupiters. The chain of logic begins with the
assumption that the system begins very well-aligned ($\psi\approx
0\arcdeg$).  A natural prediction of in-situ formation theories, or
theories involving migration due to torques from the protoplanetary
disk, is that the orbit of the planet remains very well-aligned with
equatorial plane of the star.  (However, in the latter case the planet
could conceivably misalign with the protoplanetary disk, and thus the
stellar spin, depending on which resonant and secular torques dominate
the planet-disk interaction; \citealt{1984BGT,1994WH,2001LO}.)  In
contrast, a very broad $\psi$ distribution can be produced by
mechanisms involving \cite{1962K} eccentricity cycles due to a distant
companion star \citep{2007FT, 2007WMR}. A nearly isotropic $\psi$
distribution can be produced by dynamical relaxation \citep{2001PT} or
planet-planet scattering \citep{2008N}.  It is possible that the two
components in our statistical model correspond to two different
channels for migration, one that preserves the initial spin-orbit
alignment and one that randomizes spin-orbit alignment to a
significant degree.

A possible confounding factor is tidal damping. Based on the currently
observed system parameters, it is expected that tidal coplanarization
(also called ``inclination damping'') is not a major influence on
$\psi$ \citep{2005Wa}, but this is not a watertight argument. The
timescales of long-term tidal processes are poorly known.
Nevertheless, if tides raised on the star are large enough for
substantial coplanarization, then the planet is in imminent danger of
spiraling in and being engulfed \citep{2009L}.  Moreover, one would
think that tidal damping of $\psi$ by dissipation in the star should
be \emph{fastest} for systems with the most massive planets (see,
e.g., \citealt{2007FJG,2008Jackson}). Hence it is intriguing that
XO-3b, the most massive transiting planet with an RM measurement, and
thus the one for which tidal dissipation should have been the most
important, is the only system showing evidence for
misalignment.\footnote{There is also dissipation within the planet,
  which is less efficient for higher mass planets. However, this mode
  of dissipation is less relevant because the torque on the planetary
  bulge does not couple strongly to the stellar obliquity $\psi$
  \citep{2007FJG}.} Therefore the misalignment of XO-3b suggests that
tidal damping is \emph{not} responsible for low $\psi$ values, and
that the observed low $|\lambda|$ values should be interpreted as a
relic of the planet formation and migration processes.

In this paper we have been concerned with constraining the parameters
of simple and fairly generic mathematical models for the distribution
of spin-orbit angles. A priority for future work is to use the
Bayesian framework developed in this paper to constrain the parameters
of realistic, physically motivated models, based on the specific
predictions of migration theories. In this vein we encourage theorists
to make quantitative predictions about the \emph{distribution} of
$\psi$. After deriving a theoretical distribution for $\psi$ by
whatever means, one may find the corresponding $\lambda$ distribution
for edge-on orbits using
equation~(\ref{eqn:probdensity-lambda-given-psi-edgeon}). This
requires a convolution similar to that of equation~(\ref{eq:pfisher}).
The theoretical predictions may then be directly compared with the
data. With RM measurements of sufficient quantity and quality and with
theories of sufficient specificity, it may be possible to rule out
certain migration theories, or to derive the likely fraction of
systems that migrated through different channels.

\acknowledgements We thank Ed Turner for helpful conversations about
astrostatistics, and Scott Tremaine for comments on the manuscript. We
thank the referee, Frederic Pont, for suggesting we quantitatively
analyze non-Gaussian and catastrophic errors, for the XO-3 system in
particular.  D.F.\ gratefully acknowledges support from the Michelson
Fellowship, supported by the National Aeronautics and Space
Administration and administered by the Michelson Science Center. This
work was partly supported by a grant from the NASA Keck PI Data
Analysis Fund (JPL 1326712), and from the NASA Origins program
(NNX09AD36G).

\bibliography{ms} \bibliographystyle{apj}

\end{document}